# Assured Autonomy:
# Path Toward Living With Autonomous Systems We Can Trust

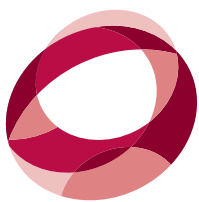

**CCC**
Computing Community Consortium
Catalyst

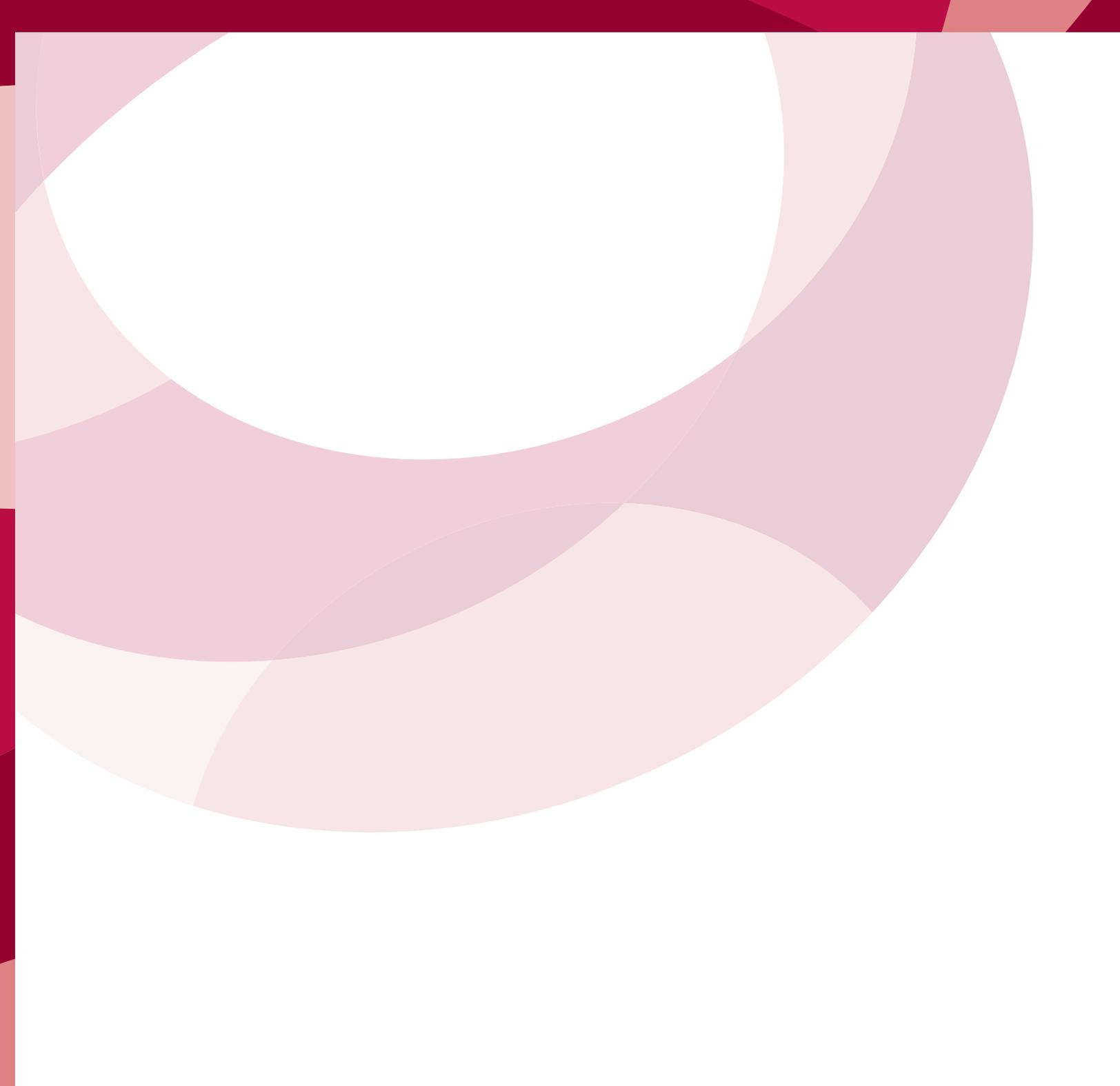

The material is based upon work supported by the National Science Foundation under Grant No. 1734706. Any opinions, findings, and conclusions or recommendations expressed in this material are those of the authors and do not necessarily reflect the views of the National Science Foundation.

# Assured Autonomy:
# Path Toward Living With
# Autonomous Systems We Can Trust

**Workshop Organizers**

Ufuk Topcu, Chair (The University of Texas at Austin)

Nadya Bliss (Arizona State University and CCC); Nancy Cooke (Arizona State University); Missy Cummings (Duke University); Ashley Llorens (Johns Hopkins University, Applied Physics Laboratory); Howard Shrobe (Massachusetts Institute of Technology); Lenore Zuck (University of Illinois at Chicago)

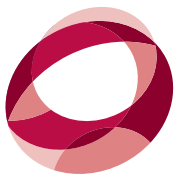

CCC
Computing Community Consortium
Catalyst

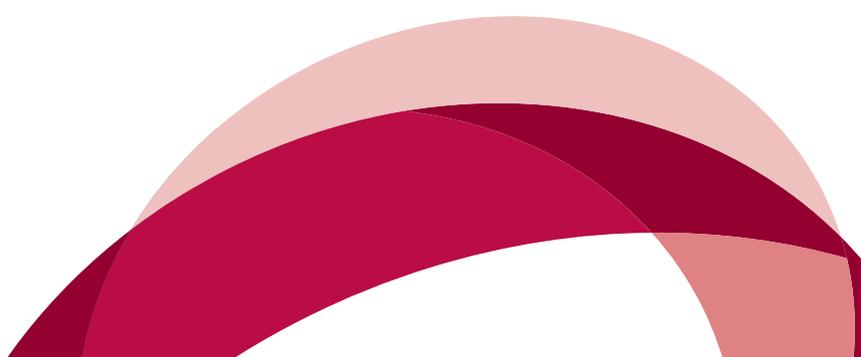

# I. Introduction

## The future has arrived – what do we want the future to be?

Science and technology create new options and opportunities for society. The printing press democratized knowledge, the industrial revolution replaced manual labor with machines, and the internet connected everyone. Each advanced society, each also disrupted society in the advance. This new age of autonomous systems portends replacing human intelligence with machine intelligence everywhere human intelligence and judgment are used. It, too, can be expected to profoundly advance and disrupt society. However, the path there is not free and not straightforward. It will be a whole-of-society endeavor to tame and shape our autonomous systems and assure that technology is safe, secure, fair, equitable, dependable, and accountable. How autonomous systems will be managed is at the crossing of science, technology, society, policy, and governance.

Society will respond, as societies have, through their institutions of government, professions, industries, education, and leaders. Throughout history it has never been a smooth process to balance the public interests, acceptable risks, norms, national security, and industry's profit motive. Public-private partnerships that nurture innovation can try the patience of governments, their stakeholders, and the public when the supply chain is too slow to respond and when risks turn into liabilities.

Autonomous systems will continue to face the same societal pressures as technology revolutions of the past have faced. However, these systems are different from earlier technologies. Replacing humans and human intelligence with machines and machine intelligence is replacing within existing frameworks of laws, ethics, morality, norms, as well as an existing technology base. A software flaw in an autonomous system, for example, is no longer just a bug; rather, it is a potential violation of law and morality.

Assured autonomy, the subject of this workshop series, is central to the discussion of the future of autonomous systems and the revolution they will surely bring. Lack of assurance manifests in longer development time, at more cost, more patching post deliverable, the acceptance of new and often unjustified risks from new sources of deployed autonomous systems, unavailability and unaffordability of insurance, and the lost opportunity costs from an important technology not existing or not available when needed most. The current pandemic offers examples of lost opportunities. With irony, the perceived benefits of autonomy, if attempted without the necessary assurance, may be negated by the lack of the assurance. Assured autonomy is about understanding and mitigating risks of operating autonomous systems in our society.

The science underpinning assurance is itself still an active interdisciplinary research area, as this workshop has shown, with critical implications on the future of autonomous systems and the society in general. Too strong a requirement for assurance with the enforcement mechanisms of government may stifle innovation while keeping the public safe today but exposing society to future shocks. Too weak a requirement for assurance places the public at risk today with little protection from future shocks.

## The Assured Autonomy Workshop Series

The challenge of establishing assurance in autonomy is rapidly attracting increasing interest in the industry, government, and academia. Autonomy is a broad and expansive capability that enables systems to behave without direct control by a human operator. To that end, it is expected to be present in a wide variety of systems and applications. A vast range of industrial sectors, including (but by no means limited to) defense, mobility, health care, manufacturing, and civilian infrastructure, are embracing the opportunities in autonomy yet face the similar barriers toward establishing the necessary level of assurance sooner or later. Numerous government agencies are poised to tackle the challenges in assured autonomy.

It is important to note that how we analyze, address, measure, and deploy autonomous systems is embedded in an existing set of laws, norms, policies, and societal values, as well as in disciplines. Controls, computer science, machine learning, artificial intelligence, human factors, communication, perceptual and cognitive sciences, business, law, ethics, and public policy are only a few among the disciplines that address the underlying scientific problems in assured autonomy.

Given the already immense interest and investment in autonomy, a series of workshops on Assured Autonomy was convened to facilitate dialogs and increase awareness among the stakeholders in the academia, industry, and government. This series of three workshops aimed to help create a unified understanding of the goals for assured autonomy, the research trends and needs, and a strategy that will facilitate sustained progress in autonomy.

The first workshop, held in October 2019, focused on current and anticipated challenges and problems in assuring autonomous systems within and across applications and sectors. The second workshop held in February 2020, focused on existing capabilities, current research, and research trends that could address the challenges and problems identified in workshop 1. The third event was dedicated to a discussion of a draft of the major findings from the previous two workshops and the recommendations.

The next two sections of the report outline the findings and recommendations. We intend these findings and recommendations to be usable by the government agencies for building and refining research and development programs and the science and by technology policymakers for setting priorities. The following two sections provide more details on the opportunities and challenges in several disciplines and applications.

## Overview of the major findings

The major advances in the technology for autonomy over the past decade or so promise a transformation driven by autonomy over the coming decades. We are at the beginning of the transformation. The findings of this workshop series outline the challenges and shortcomings that need to be addressed so that autonomy can sustainably fuel a long-lasting transformation.

The findings embrace the (already vast and likely underestimated) extent and heterogeneity of the challenges to be overcome. They also advocate for coordination among all disciplines — technical and "non-technical" — and all stakeholders.

- No good can come from autonomy without proper assurance.
- The goal of assured autonomy is human-centered — to amplify, augment, enhance, and empower humans.
- Autonomous systems have a diverse set of vulnerabilities.
- Open operation environments amplify technical challenges.
- Assurance is context-dependent and not once-and-for-all.
- Assurance for autonomous systems requires a major re-think.
- Means for measuring the progress and characterizing the gaps are necessary.
- Autonomy is a socio-economic opportunity and challenge.
- The public will perceive and be affected by autonomy unevenly.
- The challenges in autonomy require interdisciplinary approaches.
- Strengthening connections between the stakeholders will accelerate progress.
- A re-envisioning of education and workforce development must be part of the path forward.
- A national research strategy for assurance is needed.

Assurance, in this context of autonomy, cannot be an afterthought, and "assured" is not just one of the qualifications for autonomy. Autonomy can survive as a useful technology and fuel the envisioned transformation only with proper assurances. The disciplined principles and practices—yet to be developed—for assurance will create the medium in which the interdisciplinary approach this report advocates can be realized. Finally, while the emphasis should be on generalizable principles and practices, assurance will look different for different stakeholders, as each application will carry different risks and trade-offs.

## Overview of the recommendations

The recommendations aim to outline the high-level characteristics of an approach toward addressing the technical, economical, and societal challenges for the adoption of autonomy at scale and the creation of a long-term transformation fueled by autonomy. They also identify the potential roles for different stakeholders in the implementation of this approach.

We emphasize that this report intends to be an invitation — not a final statement — for a comprehensive effort that is interdisciplinary beyond all past efforts on autonomy.

## 2. Major findings

The following will distill several major findings from the workshop series. These findings point to challenges and opportunities (at times simultaneously) in autonomy. They include overarching properties that appear across disciplines and the need for change in how we do research, educate students, and/or operate autonomous systems.

### No good can come from autonomy without proper assurance

Our ability to establish appropriate assurance for autonomous systems will determine how effectively we can manage their risks and benefits. Lack of appropriate assurance may drive the cost of development uneconomically high, delay adoption of autonomous systems at scale, or result in catastrophic consequences. Lack of assurance can potentially negate the perceived benefits of autonomy.

### Human-centered autonomy

The goal of assured autonomy is to develop systems that are not only trustworthy, but that will also amplify, augment, enhance, and empower humans. Such human-centered autonomy can be achieved only by accounting for human-autonomy integration and human values throughout the design and lifetime of autonomous systems and developing (human) participatory approaches for the specification, assessment, and governance of these systems.

### Autonomous systems have a diverse set of vulnerabilities

The development of autonomous systems relies on an uncontrollable, deep supply chain and open-source software. Consequently, these systems inherit the vulnerabilities of their building blocks and potentially introduce new vulnerabilities due to the integration of heterogeneous functionality.

### Open operation environments amplify technical challenges

Autonomous systems are to operate in environments that cannot be confined within clear boundaries. By the very nature of autonomy, these environments will evolve over time, may change abruptly, drastically, and unexpectedly, and may include adversaries. These characteristics amplify the challenges in every element of assurance, from actionable specifications of behavior to verification to certification.

### Assurance is context-dependent and not once and for all

An assurance argument is useful only in its intended context, i.e., the environment characteristics, the assumptions in reasoning, the fault and threat models, etc. Therefore, it is critical to explicitly and unambiguously express what the outcome of assurance does and does not guarantee and how robust or sensitive it is to shifts in the context. Additionally, it is desirable for an assurance argument to continually adapt to the changes in the context in a manner transparent to all stakeholders.

### Assurance for autonomous systems requires a major re-think

Autonomous systems feature a level of complexity that is unprecedented in systems for which there exist principles and practices for systematically establishing assurance. This demands a major re-think of assurance — rather than trying to merely re-purpose the existing practices.

### Measuring the progress and characterizing the gaps

The breadth and interdependencies of the challenges complicate measuring the progress and raise a need for new benchmarks and testbeds. Furthermore, a catalog of the existing capabilities and acceptable reference systems may help prioritize for more pressing needs and adjust the performance expectations.

## Autonomy is a socio-economic opportunity and challenge

Autonomy will potentially affect virtually all aspects of society. Some effects are immediate, such as removing the driver from vehicles; some effects are indirect and realized over time. For example, the implications of driverless vehicles can be traced all the way to the new business opportunities, the need for new employment opportunities, the changes in the future of work, urban planning, law, governance, and policy, and the implications for the environment. A holistic perspective that embraces all socio-economic aspects — coupled with the technical aspects — of autonomy is necessary for proper incentives for assurance and adoption of autonomy at scale.

## The public will perceive and be affected by autonomy unevenly

What is an acceptable level of assurance for autonomous systems will evolve over time as people begin to understand the benefits and the risks. The impact of autonomy on different stakeholders and, even more importantly, on different demographic groups of the society will be different.

## The challenges in autonomy require interdisciplinary approaches

The challenges in autonomy are interdependent. For example, such interdependencies induce correlations and trade-offs between the technical requirements and measures for safety, security, and privacy — three of the properties autonomous systems are expected to respect. Deferring the progress for any of them may undermine the progress for the others. Similarly, coordination toward addressing the technical, regulatory, legal, and ethical challenges may create not only an impact greater than otherwise possible but also cross-fertilization.

## Strengthening connections between the stakeholders will accelerate progress

Building stronger connections between the stakeholders — academia, the industry, the regulators, the policymakers, and the public — would improve efficient use of the resources and better align objectives and approaches.

## Re-envisioning education and workforce development

An important — if not the most important — need for affordably developing and deploying autonomous systems is an appropriately trained workforce. Arguably no academic program offers the necessary interdisciplinary approach, and recent studies point to a growing mismatch between how academia perceives the gaps in education and training and how industry and government do.

## A national research strategy for assurance is needed

Assured autonomy is a national challenge. Yet there currently exists no coordinated approach, no organized community response, and no national research program searching for answers.

## 3. Recommendations

We stand at the beginning of the next wave of technological change that assured autonomy will bring. It will fundamentally alter our relationships to our critical infrastructures, our national defense, our lands, our lives, and livelihoods. In these different contexts assured autonomy must be viewed. Each context has its own risks and priorities, its own knowns and unknowns, and its expectations of consequences. Research and development for assured autonomy in the complex fabric of society is inherently interdisciplinary, convergent, crosscutting, and broad. The recommendations presented here offer a vision for exploration, innovation, and advancement that will make our society more prepared for this new technological transformation.

The recommendations build a vision for progress around a "network of institutes on autonomy" and briefly describe possible roles for the researchers, educators,

# MAJOR FINDINGS

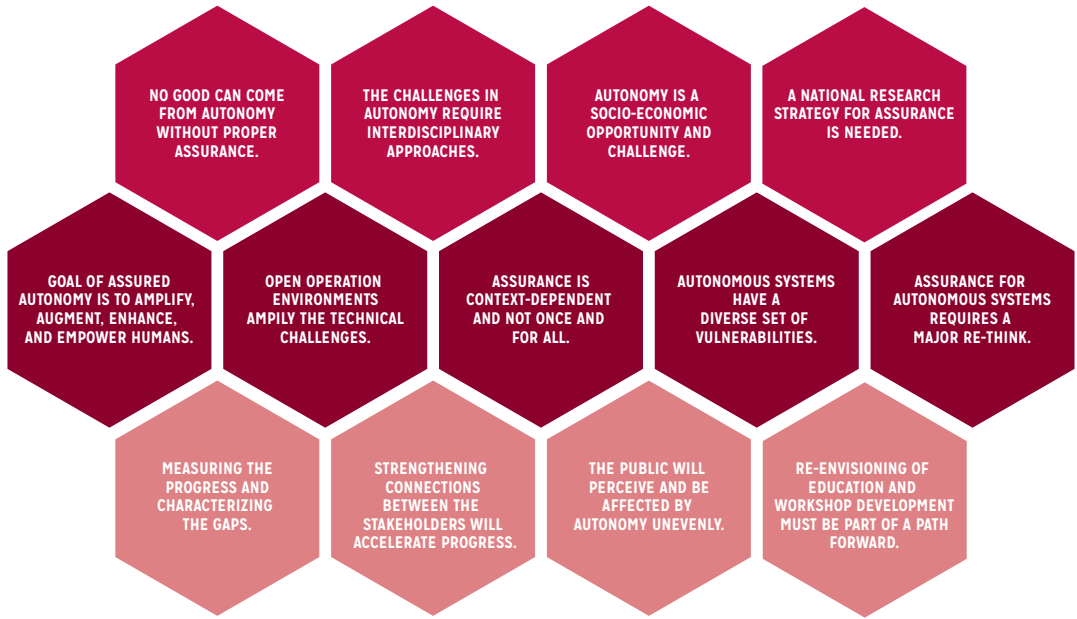

- NO GOOD CAN COME FROM AUTONOMY WITHOUT PROPER ASSURANCE.
- THE CHALLENGES IN AUTONOMY REQUIRE INTERDISCIPLINARY APPROACHES.
- AUTONOMY IS A SOCIO-ECONOMIC OPPORTUNITY AND CHALLENGE.
- A NATIONAL RESEARCH STRATEGY FOR ASSURANCE IS NEEDED.
- GOAL OF ASSURED AUTONOMY IS TO AMPLIFY, AUGMENT, ENHANCE, AND EMPOWER HUMANS.
- OPEN OPERATION ENVIRONMENTS AMPLIFY THE TECHNICAL CHALLENGES.
- ASSURANCE IS CONTEXT-DEPENDENT AND NOT ONCE AND FOR ALL.
- AUTONOMOUS SYSTEMS HAVE A DIVERSE SET OF VULNERABILITIES.
- ASSURANCE FOR AUTONOMOUS SYSTEMS REQUIRES A MAJOR RE-THINK.
- MEASURING THE PROGRESS AND CHARACTERIZING THE GAPS.
- STRENGTHENING CONNECTIONS BETWEEN THE STAKEHOLDERS WILL ACCELERATE PROGRESS.
- THE PUBLIC WILL PERCEIVE AND BE AFFECTED BY AUTONOMY UNEVENLY.
- RE-ENVISIONING OF EDUCATION AND WORKSHOP DEVELOPMENT MUST BE PART OF A PATH FORWARD.

# RECOMMENDATIONS

A vision to prepare our society for a technological transformation enabled by assured autonomy.

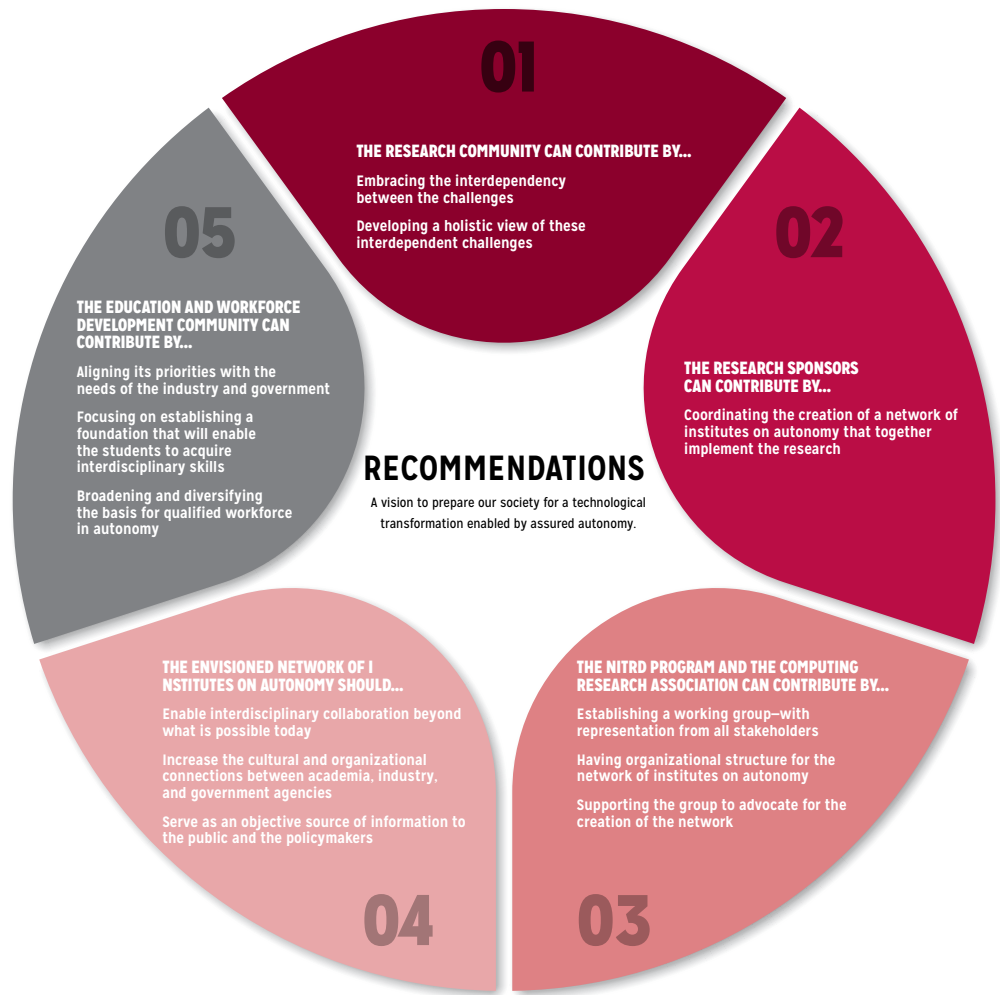

**01 THE RESEARCH COMMUNITY CAN CONTRIBUTE BY...**
- Embracing the interdependency between the challenges
- Developing a holistic view of these interdependent challenges

**02 THE RESEARCH SPONSORS CAN CONTRIBUTE BY...**
- Coordinating the creation of a network of institutes on autonomy that together implement the research

**03 THE NITRD PROGRAM AND THE COMPUTING RESEARCH ASSOCIATION CAN CONTRIBUTE BY...**
- Establishing a working group—with representation from all stakeholders
- Having organizational structure for the network of institutes on autonomy
- Supporting the group to advocate for the creation of the network

**04 THE ENVISIONED NETWORK OF INSTITUTES ON AUTONOMY SHOULD...**
- Enable interdisciplinary collaboration beyond what is possible today
- Increase the cultural and organizational connections between academia, industry, and government agencies
- Serve as an objective source of information to the public and the policymakers

**05 THE EDUCATION AND WORKFORCE DEVELOPMENT COMMUNITY CAN CONTRIBUTE BY...**
- Aligning its priorities with the needs of the industry and government
- Focusing on establishing a foundation that will enable the students to acquire interdisciplinary skills
- Broadening and diversifying the basis for qualified workforce in autonomy

research sponsors, and the organizations sponsoring this workshop series.

**The research community can contribute by…**

◗ embracing the interdependency between the challenges; and

◗ developing a holistic view of these interdependent challenges.

**The education and workforce development community can contribute by…**

◗ aligning its priorities with the needs of the industry and government;

◗ focusing on establishing a foundation that will enable the students to acquire interdisciplinary skills; and

◗ broadening and diversifying the basis for qualified workforce in autonomy.

**The research sponsors can contribute by…**

◗ coordinating the creation of a network of institutes on autonomy that together implement the research.

**The envisioned network of institutes on autonomy should…**

◗ enable interdisciplinary collaboration beyond what is possible today, including the disciplines that are currently considered to be peripheral;

◗ Increase the cultural and organizational connections between the academia, industry, and government agencies; and

◗ serve as an objective source of information to the public and the policymakers.[1]

**The NITRD Program and the Computing Research Association—sponsors of the workshop series—can contribute by…**

◗ establishing a working group—with representation from all stakeholders—to make concrete the vision, working principles, and organizational structure for the network of institutes on autonomy; and

◗ supporting the group to advocate for the creation of this network.

## 4. The challenges, trends, and needs

This section details the challenges, trends, and needs that were brought up during the workshop panels and breakout discussions and in the focused virtual panels conducted after the workshops. While this overview breaks the presentation into several categories, there is a tight coupling among the challenges, trends, and needs in these categories.

### 4.1 Safety and verification

Over the last several decades, the field of formal methods for verification and testing has amassed a rich body of principles and techniques that can be used to improve the reliability and safety of engineered systems. Autonomous systems have unique features that have limited the impact of formal methods on assurance. For example, autonomous systems often execute in environments that are partially unknown and can change unpredictably, and these environments can pose safety hazards. Additionally, autonomous systems tend to incorporate learning-enabled functionality and run the risk of learning unsafe behavior or creating excessively complex artifacts. They also typically must share decision-making responsibilities with human operators and hand off control to a human at specific points. However, coordination between the system and its operator may introduce additional sources of failures with safety-critical consequences. These relatively unique characteristics necessitate new forms of requirement specification, new abstractions, and new algorithmic verification techniques.

The lack of actionable and unambiguous requirements for new categories of autonomous systems is a significant bottleneck for verification. Such requirements—analogous to datasets in machine learning—and evaluation paradigms

---

[1] https://www.state.gov/declaration-of-the-united-states-of-america-and-the-united-kingdom-of-great-britain-and-northern-ireland-on-cooperation-in-artificial-intelligence-research-and-development-a-shared-vision-for-driving/

developed and agreed on by the research community, have the potential to drive the creation of new research techniques and serve as an impetus for practitioners and regulators to adopt research from the community. For example, presumably autonomous systems should be required to be robust in conditions of environmental uncertainty and take corrective actions in response to unforeseen changes that endanger safety. Additionally, the ability to trace the root cause of failures is critical in autonomy. Determinism—that the system performs the same action at a given state no matter the context—is a requirement that facilitates such tracing. On the flip side, certain circumstances may dictate a requirement to take calculated and adjustable risks for higher performance.

In addition to requirements, the community should develop abstractions that facilitate the principled design of autonomous systems and facilitate the static or dynamic enforcement of requirements. In particular, such abstractions should contain the influence of the system components that are difficult to verify—for example, neural network used for learning—and rigorously vet the outputs of such unverified components. In addition, users need a mental model of why the system is making the decisions it is making so that they can judge when to override the system's actions. Principled abstractions can go a long way in ensuring a seamless interface between the system and the human operator.

Autonomous systems are a natural candidate for algorithmic techniques, including model checking, abstract interpretation, deductive verification, and run-time enforcement. However, because autonomous systems are typically heterogeneous, their verification is likely to require a range of methods rather than a single unifying tool. For example, it may be possible to formally verify the control module of an autonomous system; however, verification of perception modules implemented as neural networks would necessarily be statistical. Separately, the community should invest in methods to compose module-level guarantees produced by different tools into end-to- end guarantees. The outcomes of these techniques should be interpretable by the designers and regulators; provide guidance to reduce the necessary testing effort; and help eliminate unintended functionality; and identify the gaps in the requirements.

Perhaps the most remarkable difference between conventional safety-critical systems and autonomous systems is that, in the latter, learning and adaptation play a central role. That is, the knowledge about the system and the environment that is assumed in formal verification changes at run time, and static, design-time verification is likely to be insufficient. The approach must fundamentally shift from "design, verify, and deploy" to "design, provisionally verify, continually re-validate the assumptions, and continually iterate the design."

Therefore, the community should invest in methods that are adaptive and tightly coupled to learning and verification and receptive to the computational, power, and thermal limitations at run time.

### 4.2 Security

A secure system is one that will behave as designed and implemented even when under attack. It follows that security is a necessary condition for assured autonomy. Each year thousands of conventional computing systems have been subverted by cyber-attackers. Some well-known examples include the Stuxnet attack that corrupted the software at a nuclear enrichment plant in Iran in 2010, a cyber-attack on the control system for power distribution in Ukraine in 2015, and the cyber-attack called Car Shark that took control of a conventional car in 2010. No autonomous system can be trusted to behave as intended if it is vulnerable to cyber-attack.

A key enabler of cyber-attacks on autonomous systems is network access allowing attackers to penetrate systems remotely from anywhere in the world.[2] Other potential attack vectors come from input to the system, sensors in a computer vision system, and/or audio input. Most autonomous systems are networked partly to allow for remote maintenance, monitoring, and upgrades, but increasingly network connectivity is inherent to their design and operation. Additionally, underlying the highest and more visible layers of autonomous functionality, there is a deep stack of conventional software: operating systems, run-time systems, and application-specific libraries that are crucial to the correct operation of the autonomous system. These large layers of software, developed over a long period, are known to contain cyber vulnerabilities.

---

[2] https://ec.europa.eu/commission/presscorner/detail/en/IP_19_6049

In addition to vulnerabilities emerging from reliance on conventional network and software layers, autonomous systems have several unique requirements that introduce new sources of vulnerabilities in the autonomy layers: (i) Autonomous systems must perceive the world accurately. This need typically requires the systems to receive unmodified signals from a variety of sensors; they must fuse these signals, and then construct a model of what they are perceiving. (ii) They must control the physical system by issuing commands to their effectors, and these commands must arrive in unmodified form. (iii) The delay in the decision cycle (including sensing, interpreting, and acting) must meet strict timing requirements. (iv) The resource (e.g., sensing, processing, communication, and power) limitations introduce requirements competing with the system's ability to defend against attacks, particularly in open and possibly contested environments.

One of the more pressing challenges stems from the use of modern perceptual techniques that increasingly rely on data-driven machine learning technology with mathematical constructs (e.g., deep neural networks) that are not aligned with conventional abstractions and behavioral specifications. For example, these systems are used to train "classifiers" that recognize objects within images. They have already potentially played roles in publicized autonomous vehicles incidents and are known to be vulnerable to adversarial patterns, misleading examples, and data poisoning.

The first of the problems comes from the reliance of the machine-learning-based frameworks on large image processing libraries; several researchers discovered cyber vulnerabilities in these libraries. Attackers can use these vulnerabilities to modify incoming sensor data as well as to divert the control flow of the system to attacker-provided code. The second issue arises from the dependence on large training data sets. Attackers have found at least two ways to corrupt systems that rely on these large data sets: "poisoning" the training data to cause the system to be incorrectly trained; and using "adversarial examples" or images designed to fool the classifier.

Until these cyber-security issues are resolved, there cannot be much faith that an autonomous system will behave as intended. Sadly, the advantage is still with the attacker in an "arms race" in which defenders discover a vulnerability and patch it while the attacker modifies existing attacks in ways that bypass the defense. Attackers seem to be able to innovate faster than defenders, allowing them to maintain an advantage. There is an ongoing supply chain. Suppliers from many locations increase the attack surface by orders of magnitude. Attackers merely find one vulnerability to exploit while the defender must defend against many vulnerabilities.

Changing the imbalance demands fixing the way the systems are designed and tested and how software is verified. Because autonomous systems operate in an open world, traditional testing and verification techniques need to be extended to deal with the fact that not every eventuality can be predicted. The necessary change can be achieved only through an interdisciplinary approach that integrates insights from all relevant disciplines throughout the design flow and lifetime of the systems.

## 4.3 Certification and regulation

Certification and regulations for autonomous systems are, at best, in their infancy. They have a role in determining the timeline for the deployment of autonomous systems, protecting the public's safety and interests, creating incentives for assurance, and facilitating the creation of an ecosystem that promotes progress in businesses related to autonomy. Assurance in autonomy (or in any other emerging, safety-critical application) faces a chicken-or-egg dilemma: systems cannot be deployed at scale without proper assurance, but there can be no "complete" assurance without deployment at scale. Certification and regulations have the potential to help resolve such a dilemma through calculated risk-taking by accounting all aspects of assurance.

Certification is a complicated and costly process even for relatively conventional safety-critical systems, e.g., civil transport aircraft and nuclear plants, for which the operating environment can be restricted tightly with humans handling exceptional situations. The blurred lines separating system and human authority and responsibility in an open operation environment complicates the certification of autonomous systems. Other problematic factors include the lack of continuous human oversight, the need for transparency for a human to react, the reliance on continually updated functionality (due to integration of learning-enabled elements), the potential exposure to non-cooperative actors, and emerging privacy concerns. Certification for autonomy will require a major re-thinking as even how one can establish specifications against which the systems should be certified—let alone the certification process itself—is currently unclear and underexplored. Even assured autonomy would not remove the need for human operators to be licensed in some way.

The regulatory and legal system will need to address issues of responsibility and liability with regard to autonomy. Even in systems overseen by human operators, autonomy may already be the cause of faults, for example by providing erroneous, insufficient, or late feedback on problems occurring in the system. The questions that need to be addressed include the following:

◗ Who is responsible when a fully autonomous system causes a loss event?

◗ Who is responsible when an autonomous system interacting with humans causes a loss event when it is not evident whether fault lies with the human, the autonomous system, or a mix of both?

◗ If liability waivers are made available to manufacturers and/or operators, will that reduce pressure to create safe vehicles, resulting in "gaming" of safety standard compliance?

The potential reasons to regulate autonomous systems include: aligning industry incentives with the public interest by preventing immature technology that causes potentially unacceptable risks; increasing transparency even if compliance with standards is not required; increasing transparency of the basis for decision outcomes; and mitigating potential bias. On the other hand, the regulations should not create a barrier for entry for smaller companies that offer technical value but lack the resources to engage in an onerous certification process.

Maturation of certification processes and regulations for autonomy can benefit from: developing design approaches that treat testing, verification, and certification as first-class objectives rather than afterthoughts; focusing on transparent and formal techniques with explicit feedback on the gaps in coverage (as emphasized the standards UL 4600, ASTM F3269-17 and AAMI 2700-2-1); understanding the compromise between regulation and flexibility; standardizing desirable properties at a high level of abstraction rather than at the level of implementations; harmonizing the standards and the liability and regulatory approaches internationally; and improving the education of regulators about the technology.

### 4.4 Human-system integration and trust

The goal of the future progress in autonomy should be to have human-centered interdependence autonomy that is not only reliable, safe, responsible, and trustworthy, but that will also amplify, augment, enhance, and empower people. In fact, the best human-autonomy team should be able to outperform the best human or the best machine. Alternatively, if the autonomy does not behave as humans would expect, or if it is unreliable, or if it lacks transparency, then humans will not use it, will mistrust it, or will over-trust it. The literature and recent incidents witness the potentially catastrophic consequences of disconnect from humans.

Human-centered autonomy can be achieved only by considering human-system integration needs throughout the system's lifetime. Sample questions that guide various stages of the lifetime include the following:

◗ In the specification stage, how do we translate our goals for human society into performance specifications for autonomous systems?

◗ How is contextual knowledge transferred from humans to the machine during the design process?

◗ What test and evaluation paradigms are needed to understand the contexts in which systems will perform safely and effectively?

◗ In system operation, how do we translate contextual knowledge of system capabilities to human operators (particularly when they may not be versed in the underlying technology)?

◗ During system evolution, when do system updates require recertification of the system, the human operators, or their combination, and when do they need human operators to be retrained?

The recent trends toward human-autonomy teaming offer a promising medium to investigate human-system integration needs. As Johnson and Vera[3] articulated, "No autonomy is an island." Therefore, assurance of autonomy will need to be interpreted in the context of a team with humans and other autonomous agents. Autonomous agents in this setting are interdependent with (rather than independent from) the rest of the team and will require teamwork, communication, and coordination skills. Autonomy can also be a better team player with improved transparency, self-awareness (understanding of limits), and appreciation of teamwork. Despite tremendous progress in this topic area, significant challenges persist, such as system and information architectures that promote transparency, bidirectional transparency, real-time and dynamic generation of transparency content, effects of individual and cultural differences, and emerging unintended consequences due to the dependencies between the human and autonomous functionality.

Human-autonomy integration cannot be considered as a last resort. It is essential to account for human values by making them explicit through stated assumptions, operationalizing those values with engineering requirements from the early stages of the design process, and creating interdisciplinary approaches and teams that can facilitate early integration.. An additional requirement is developing standards for the assurance of autonomous systems that do not incur additional cost or financial barriers to design account for human values.[4]

Despite recent progress in the use of augmented reality, a lack of testbeds and methodologies for measuring the human-autonomy team effectiveness in settings that resemble the complexity of the intended operation environments plays a role in the widening gap between technology development and the deployment of autonomous systems with appropriate assurance.

## 4.5 Ethics

Technology has implicit, and sometimes explicit, embedded values, despite a common tendency to consider technology as fundamentally value-neutral. Assurance of autonomy and whether humans will trust autonomous systems likely hinge on—in addition to the other aspects discussed in this report—whether autonomous systems will operate in accordance with such embedded values without human intervention. Autonomous systems can have profound impacts on people and communities. They have the potential to reduce inequities and create positive social impact. Conversely, they can also amplify and codify adverse social outcomes.

While the meaning and role of ethics in the context of autonomy is still debated, ethics may be interpreted as a means for a society to encourage its members to behave in cooperative ways that benefit the society.[5] It is, therefore, a fundamental component of assured autonomy to encode (some) human values, resolve the conflicts in the interactions with the technology, influence the humans' perception of the trustworthiness of autonomous systems, and create a desired social impact.

Recent years have seen a growing realization of the negative consequences of autonomy, other emerging technologies, and even those that are regarded as mature (e.g., the social media platforms) when ethics are considered as an afterthought in the design. Instead, ethics needs to be designed into technology from the outset. A proliferation of principles for the ethical design of autonomous systems from has emerged from governments, nonprofits, and industry. These principles—often fragmented and typically applying to the role of artificial intelligence—focus on elements such as fairness, transparency, reliability, accountability, and security, leaving a wide-open area for developing integrated approaches accounting for all contributing factors and their interdependencies.[6]

---

[5] https://unsw.adfa.edu.au/defining-ethics-and-law-autonomous-systems

[3] https://aaai.org/ojs/index.php/aimagazine/article/view/2842

[4] An example of such a framework is found in the report Ethics Guidelines for Trustworthy AI (1), which begins with a commitment to the common good and the goal of improving human welfare and freedom, leads to four ethical principles, and, in turn, leads to seven key requirements related to human values.

Improved lifecycle tools for assurance as well as a robust interdisciplinary research agenda that engages with ethicists in the humanities, social sciences and policy and creates a common medium and languages can help operationalize ethics in the development and operation of autonomous systems. While the interplay between ethics and autonomy technology is at its infancy, recent efforts have attempted to create practical tools to incorporate ethics into the design, verification, implementation, and evaluation of autonomous systems.[7][8][9][10][11] Moving forward, these tools will need to account for the differences in the value-based grounding of ethical principles across contexts, cultures, customs, and jurisdictions.

With autonomous systems and their potential to replace human intelligence in jobs and technologies, the space for risks is as broad as society. We can, however, ask what values society believes must apply to living with autonomous systems. Values have the potential to steer the direction of research in autonomous systems and inform policymakers of the technology advantages and limitations.

## 4.6 Privacy

Although there is no universally accepted definition, privacy may be interpreted as the right of the individuals to control where their data are used. Technology typically makes it easier to collect data, monitor people, and make inferences about people. Autonomous systems raise new concerns or amplify the existing concerns because data collection is increasingly more pervasive, adaptive, responsive, and because autonomous systems directly interact with humans in the physical world.

Collecting data is a privacy concern in itself because, regardless of the purpose for collecting the data, others may access the data. Thus, the mere storage of it can create privacy concerns later. Additionally, the inferences that can be drawn from the data may be outside the imagination of the technology developers and users, regulators, lawmakers, and the public. Furthermore, the lack of algorithmic maturity and the mismatch between the ability to collect data and the ability to turn the data into task-oriented knowledge could result in the collection of unnecessary data.

Data collection by autonomous systems tends to be invisible and transparent. For example, monitoring by police is an activity more noticeable by humans compared to aerial surveillance. Additionally, autonomous systems have the potential to shape people's environments so that people believe they are acting in their self-interest. Yet, the system is actually intruding on their decision autonomy.

The meaning of privacy depends on social interactions and human values. It is, therefore, necessary to develop the technical means to facilitate an informed public debate, while being aware of the risks of identifying the standards and guidelines for the assurance of privacy in the development and operation of autonomous systems. These standards and guidelines should embrace the diversity among the individuals of the society and not focus on merely a hypothetical "average" individual.

Privacy is not a standalone concern; it introduces trade-offs with the other aspects of assured autonomy discussed in this report. For example, autonomous vehicles may make the roads safer by sharing data with other vehicles and the infrastructure, but this data sharing may jeopardize the privacy of the individuals. Therefore, it is vital to develop a holistic approach that can address not only privacy—or other aspects of assurance—in isolation, but also help navigate the trade space of the possibly competing requirements. Such an approach should develop and integrate models of privacy and specifications of expectations into the design of autonomous systems.

---

[6] https://blogs.asucollegeoflaw.com/lsi/2019/03/01/marchant-on-soft-law-and-artificial-intelligence/

[7] https://ethicsinaction.ieee.org/#series

[8] https://standards.ieee.org/project/7009.html

[9] https://ethicsinaction.ieee.org/#series

[10] https://www.amazon.com/Machine-Ethics-Michael-Anderson/dp/0521112354

[11] https://ieeexplore.ieee.org/document/8648363/

## 4.7 Societal impacts

Forecasting the societal impacts of autonomy is no easy task. Autonomy is a behavior, and as such, it can be present in a wide variety of platforms and in ways we cannot currently predict.

Rather than to focus on one particular manifestation of autonomy, categories can characterize values and thus the societal impacts of autonomous systems. For example, one set of categories may distinguish between the harms, wrongs, benefits, and offenses associated with autonomy:

- Harms: mental, physical, socio-economic harms to persons; environmental harms; mental or physical harms to nonhuman living entities.

- Wrongs: violations of certain legal rights.

- Benefits: the receipt of some form of good (e.g., monetary, access, opportunity, increase in living standard, or well-being).

- Offenses: neither harms nor wrongs, but cultural mores that induce particular distaste.

These impacts are not mutually exclusive. For example, a particular use of autonomous trucking, may be beneficial for some stakeholders, but harmful for others. There may be no outright wrong done; that is, no one's legal rights are violated, but there may still be disruption and harm to truck drivers, their families, the environment, or the surrounding communities. Likewise, benefits may accrue to the companies and shareholders, and to the public in the form of reduced traffic congestion or enhanced highway safety.

It is essential to understand the causal or temporal relations between harms, wrongs, benefits, and offenses and the accumulating and compounding effects. For example, it is plausible that autonomy in cars saves lives. Autonomy in this context has additional—though perhaps not as immediate—impacts on urban planning, the environment, where and how autonomous vehicles will be used, and whether there will be disproportionate effects on vulnerable populations. Depending upon the maturity of this technology, as well as the incentives to deploy, urban planners may have to re-engineer parts of the infrastructure, e.g., pedestrian crossings, sidewalks, overpasses, to ensure pedestrian safety while the cars are in operation. Yet, such re-engineering may have tertiary effects on the populations that live in these areas, creating different ways of mobility, access, and opportunity. Consequently, additional costs, monetary and otherwise, must be balanced against the primary and stated value of "saving lives" from fatal traffic accidents.

The implications for society are complex. For example, delegating essential decisions to automated systems may cause skill erosion and decay in the ability for humans to exercise their judgment meaningfully. Such conclusions then press on the moral and legal accountability for human use of such systems, and where the accountability for such systems lies. It might be said that using these systems yield the benefits of freeing up time and mental bandwidth or increasing overall productivity. It might also be said that there is something morally offensive about delegating particular tasks or decisions to autonomous systems.

Closely connected to the societal impacts of autonomous systems are impacts on the environment. Tackling climate change and its threats to societies across the globe is among the most critical areas for making human progress. A key opportunity, therefore, is to explore how autonomous systems can reduce humans' impact on the environment. In some scenarios, autonomous vehicles might reduce the use of energy and natural resources while also offering societal benefits. In other scenarios, the environmental impacts might be substantial or unknown—wide-spread consumer ownership of autonomous vehicles might lead to more trips and more energy consumption because drivers do not lose time to traffic congestion. Accordingly, when developing requirements for autonomous systems, the Earth should be a key stakeholder.

The different uses of technology for autonomy in different domains, such as in the civilian and military applications presents another complication. It is unclear to what extent the presence of autonomous systems in these domains upholds particular values or diminishes them, and whether harms, wrongs, benefits, or offenses are likely to occur in one domain or another.

The crux of the matter is that there are always multiple factors to consider, in various temporal and causal chains, in different societal domains, and for fundamentally different tasks. The question on how to measure these impacts are as much moral and political as they are technical.

Determining the causality and compounding relations between the societal effects is a challenge that can be addressed only through the comprehensive and interdisciplinary approach advocated in this report. Yet these determinations is the precursor for influencing the evolution of autonomy toward a net-positive impact.

## 4.8 Education

A recent NSF-supported survey investigated how academia may need to adapt curricula for aerospace autonomy education.[12] The goal of the survey was to determine the gaps between the deliverers of education (academia) and those customers that hire graduates (industry and government). The survey generated 132 responses distributed roughly equally across academia, industry, and government. While the primary target of the survey was aerospace autonomy, the results likely shed light on the needs and trends in education for autonomy in general. Among the significant observations from the results of the survey:

◗ About 80 to 90 percent of the respondents from industry and government identified an urgent and critical need for graduates with backgrounds in autonomy.

◗ Only 30 percent of those respondents were confident that new hires were prepared for careers in autonomy.

◗ When asked about the relative importance of various topics, a disparity emerged. While academics considered machine learning to be the most important topic, respondents from industry and government identified safety, cybersecurity, system integration, and certification as more pressing needs.

◗ The respondents from industry and government pointed to a need for workforce retraining.

This survey illustrated a dramatic need for interdisciplinary education and research in aerospace autonomy. The issues of safety, testing, and certification raised by industry and government have broad application, and not just for aerospace systems. Autonomy is becoming a core technology in a wide range of other safety-critical settings. Additionally, the needs are not specific to the United States. For example, the outcomes of a recent joint US-UK workshop[13] show significant parallels with the observations outlined above.

The student body associated with the specialties relevant for autonomy is exceptionally diverse, both in maturity and in prior expertise, and will further diversify as the field moves from proofs-of-concept for the technology to integration at scale and creation of positive societal impact. Therefore, establishing a foundation to enable the students to acquire interdisciplinary skills will be essential.

Finally, a societal transformation fueled by autonomy and the necessary transformation in education can be achieved only by expanding and diversifying the basis for the workforce that will sustain the growth in the research, development, deployment, operation, and governance of autonomous systems. To this end, it is critically important to create a culture of inclusion in education for autonomy and in the field of autonomy through education.

## 4.9 Governance and policy

Governance—and the associated policy tools — for autonomous systems make up a critical element of the interdisciplinary approach, given the far-reaching implications of long-term transformation and the diversity and extent of the challenges and opportunities.

---

[12] Cummings, M., Morgansen, K., Argrow, B., & S. Singh, "Transforming Aerospace Autonomy Education and Research" IEEE Aerospace 2021, Big Sky Montana.

[13] Held by the US-UK Science and Technology (S&T) Defense Review (US-UK S&T Stocktake) committee.

Assurance is essential for society to understand what risks it is taking with the increased adoption of autonomy. The weaknesses in our understanding of assurance for autonomy manifests itself in policy. Often it is not until a harm cannot be ignored that a policy is patched. A strategic societal and policy approach to this technology revolution is preferable to a patchwork of exceptions to redress a real (or imagined) harm.

Understanding acceptable risks and risk mitigation are critical steps to making good policy for autonomous systems and for deciding what future with autonomy we want and how to get there. With the self-driving car prototypes on the roads today, society, and particularly policymakers and technologists, must decide on the paths going forward and with what acceptable assurances. The strategies adopted and investments made will determine how well and how deeply autonomous systems become part of society.

The governing bodies, interpreted broadly beyond the federal government, have the responsibility to represent the public interest by regulating its risk appetite and correcting for the failures and timelines of the market (e.g., attempts to deploy autonomous systems without proper assurance). In principle, these governing bodies are supposed to predict the concerns related to security, privacy, fairness, accountability, and liability and take steps for preempting, mitigating, or compensating for the socially unacceptable outcomes of these concerns. Additionally, ensuring access to resources, information, and opportunities to participate in the development and operation of autonomous systems hinges on their decisions.

Benefit redistribution throughout the envisioned long-term transformation fueled by autonomous systems is also among their responsibilities. While this redistribution may seem to be an issue of the future. the impact of autonomy on the society will potentially be systemic, and failing to account for benefit redistribution after large-scale adoption of autonomous systems will only worsen the benefit imbalance.

The roles for the governing bodies include: establishing mechanisms and boundaries to prevent an unjustified shift of risks (e.g., from the commercial sector to the public) and liability (e.g., from the developers to the operators to the insurance sector or the reverse); providing third-party oversight of development and deployment; and conducting third-party review of the autonomous system market. In particular, this oversight may involve imposing algorithmic transparency (for the benefit all stakeholders, notably including the regulators and the public), certification requirements and guidelines, and standards for supply-chain interrogation and product deployment. Market overview may take the form of ensuring the timeliness of the product delivery and setting audit criteria and reporting requirements.

Making informed policy involves understanding the risks, their likelihood, and their consequences. Making decisions on accountability, individual harms, public adoption, and redress requires understanding the algorithms used in autonomous systems. The harms may not only be physical damage, but also social harms related to equality, equity, fairness, ethics, morality, and acceptable risks. Moreover, these systems will learn from their experience, making our capability to audit, report, and enable third party oversight a significant challenge. Finally, since legal and moral responsibility for failures of autonomous systems resides only in humans and organizations, designers and developers should be trained to understand the legal doctrines of liability and accountability.

Among the most pressing hurdles for the governance of and policymaking for autonomous systems is the absence of legacy examples and sufficient empirical evidence. Coupled with the lack of access to knowledge and technical expertise, it is becoming increasingly difficult to draw generalizable insights from the limited amount of and likely biased test data. This increases the likelihood of decisionmaking errors (with potentially catastrophic consequences) or of making shortsighted decisions.

If and when autonomous systems operate with assurance at near-peer levels of human performance and beyond, the relationships among people, corporations, and autonomous systems will evolve and be codified into accepted practice, profoundly affecting the future of work and living in society.

Governing principles for autonomous systems (and the neighboring field of artificial intelligence) are emerging[14][15][16] along with the disparities in the approaches, e.g., between the US and Europe on the level of restrictions. Dictating the right approach for governance and policymaking for autonomous systems is beyond the scope of this report. However, we emphasize that there is no shortcut for appropriate assurance in autonomy, and the choices that seem cost-saving today may result in higher costs in the future.

## 5. Use cases

This section presents a collection of use cases of autonomous systems. While this exposition is highly biased toward the interests of a subset of the workshop participants and is nowhere close to offering a fair coverage of the potential uses of autonomy, it is intended to provide a concrete interpretation of the challenges, trends, and needs (discussed in the previous section at a general and abstract level) in the context of a few applications. The extent of the potential uses for autonomy will be limited only by our imagination and our ability to establish appropriate assurance for what we can imagine.

### 5.1 Urban air mobility

Urban air mobility (UAM) has the potential to revolutionize mobility around densely populated metropolitan areas by providing a practical air transportation system for passengers and cargo. A convergence of evolving technologies, such as electric propulsion and increasing levels of autonomy, along with the emergence of new business models, such as mobile application-based ride- sharing, may act to spur the development of this mode of transportation and address growing mobility needs in cities. Realizing this ultimate vision will require novel aircraft designs with relatively low noise and high efficiency, combined with innovative airspace operations management systems that can safely and efficiently manage large numbers of aircraft over small areas. Furthermore, to be an effective means of transportation in metropolitan areas, UAM operations must be tightly integrated with local communities and existing modes of transportation.

Multiple technical, societal, and regulatory challenges must be overcome before the full potential of the UAM vision can be realized. The UAM operational concept relates to aspects of vehicles, airspace, and public acceptance. Furthermore, both the categories of vehicles and airspace can be further subdivided into design and development and management and operations. Autonomy spans all of these aspects: Both the vehicle and the airspace may involve increasingly autonomous systems in terms of design as well as operations, and the path to public acceptance of increasingly autonomous systems is unclear.

Specific barriers related to autonomy in UAM include:

◗ Safety: Assuring increasingly autonomous systems in a high-confidence, critical-infrastructure context is a crucial gap in current capabilities..

1) Regulatory approval (including certification, operational approval, and access to the national airspace): Requirements, standards, and corresponding means of compliance to establish the acceptability of increasingly autonomous systems from vehicles to the airspace to the UAM infrastructure – all are lacking. This lack pertains to both the design (certification) and operation (operational approval) of these increasingly autonomous systems.

◗ Security (physical and cyber): Assuring physical platforms (e.g., vehicles and critical infrastructure) security as well as cyber-security of critical communications, navigation, surveillance, information, and control infrastructure (e.g., vehicle, airspace) remains a barrier to the success of a UAM operational concept.

◗ Human-machine teaming: Novel operational paradigms, including remote operation as well as m:N operation (e.g., a small number m of possibly remote pilots flying a large number N of vehicles) currently have no path to gain operational approval to access the national airspace.

◗ Integration at scale: Operational scalability is financially feasible only with increasing autonomy in both the vehicle (e.g., remote, m:N, or autonomous operation) and the airspace (increased density of operations enabled by autonomous trajectory and fleet management). These systems must also share the airspace with legacy aircraft, operations, and infrastructure.

---

[14] https://www.congress.gov/bill/116th-congress/house-bill/2231/actions
[15] https://shop.bsigroup.com/ProductDetail?pid=000000000030320089
[16] https://www.astm.org/Standards/F3269.htm

Some of the gaps in the current capabilities are related to evidence-based safety arguments that are capable of establishing the safety of an increasingly autonomous system; analysis and simulation in a scalable fashion, to yield evidentiary artifacts that can be used to substantiate a safety claim (in whole or in part); verification and validation techniques that enable the analytical evaluation of novel features such as non-determinism, complexity, and uncertainty; standardized requirements that, when met, demonstrate that the system is safe; understanding of the trade-offs between safety and security, as well as whether techniques to establish one quality can be leveraged in the establishment of the other quality; and lack of understanding of the continuum of human-machine teaming and their roles for assurance.

## 5.2 Space

Onboard autonomy is vital for space exploration because it is often the sole solution for accomplishing highly ambitious missions. For example, the conceptual Europa Lander mission would land an un-crewed spacecraft on Europa, an icy moon of Jupiter which may harbor extraterrestrial life in its subsurface ocean. The objective of the mission is to perform a series of complex activities and search for biosignatures, including selecting the excavation site, excavating the surface to approximately 10 cm deep, obtaining at least three samples from the trench, and analyzing the samples. The lander would have only about 20 days of lifetime on the surface to complete the mission. When an anomaly occurs, waiting for the commands from the ground operation would simply be a waste of the precious lifetime. Preferably, it has to make decisions on its own to continue activities and achieve the mission before the lifetime expiration. Autonomy is crucial for future crewed missions, too. Some have estimated the level of complexity of a future crewed deep-space vehicle to be similar to that of a nuclear attack submarine, which has 134 crew members. In contrast, space missions in the next few decades will likely carry only four to six crew members.[17] Therefore, deep space human exploration missions will require autonomous vehicle systems management to maintain the operation of the complex spacecraft and detect and respond to events, such as solar flares and system failures, without crew or ground interference.

Assurance of autonomy is essential for at least two core reasons: the reliability necessary for space missions and the need for acceptance of autonomy by humans. Software or hardware failures caused by malfunctioning autonomy could immediately result in a mission failure. (Engineers cannot go to Mars to fix rovers.) In the case of crewed spacecraft, the lives of astronauts would be at stake. Essential properties that space-bound autonomy algorithms must possess are: robustness; fault detection, isolation, and recovery; graceful degradation; and verifiability.

Furthermore, like many other emerging technologies, autonomy faces healthy skepticism when introduced to flight projects. To overcome the skepticism and get adopted by space missions, autonomy has to gain the trust of humans. In other words, it is not sufficient to make autonomy functionally assured; it has to provide a sense of assurance to its users to get accepted by and contribute to future space missions.

Many of the technical challenges discussed in the previous section find their interpretations in space applications. The verification and validation of modern autonomy algorithms are becoming increasingly challenging because, as the number of internal states grows exponentially, safety cannot be assured for every possible state within a reasonable time and resource using the conventional means. Military missions would be highly sensitive to the security of the onboard software as well as the ground systems. Certification of spacecraft autonomy would be increasingly complicated because of the quickly expanding operational envelope, reduced interaction with a ground operation, and substantially increased level of complexity in autonomous behaviors. Gaining the trust of humans is important for space-bound autonomy, yet very challenging because the stakes are very high while in-space demonstration opportunities are scarce. Future space missions will not be able to harness the benefits of the quickly evolving autonomy technologies unless these challenges in assurance are overcome. Therefore, investment in the research and development efforts in assured autonomy is indeed a key for changing the culture of aversion from autonomy in space applications and enabling the human race to *boldly go where no one has gone before*.

---

[17] https://www.nasa.gov/sites/default/files/atoms/files/strategic_space_technology_investment_plan_508.pdf

## 5.3 Smart cities and response to systemic disruptions

As urban population density increases, urban infrastructure and services become stressed and unable to operate efficiently. To accommodate the influx of people, cities have made efforts to become "smarter" by employing sensors, networks, cyber-physical systems, and artificial intelligence to facilitate urban operations. The ecosystem of these technologies is the hallmark of a "smart city." The application areas of such smart cities abound, from enabling smoother traffic patterns to improving safety through increased surveillance capabilities. Smart cities have increasingly employed autonomous systems (or systems with high levels of automation) to serve different purposes. Assurance is critical for such systems where provisions are needed to guarantee both safety and security. The smart city use case of pandemic management is one such example where assured autonomous systems are required.

The COVID-19 pandemic has frozen major urban centers, requiring restrictions on activities that society previously took for granted, such as eating in restaurants. Re-purposing the autonomous systems that are already part of smart city ecosystems could serve to manage pandemic response in cities if accomplished safely and securely. Specifically, systems that have historically enabled mobility services and public safety can be re-imagined to dynamically control our surrounding environment and facilitate our safe traverse of cities.

Grocery shopping has proven to be a potentially dangerous activity for high-risk groups of people, leading to increased use of grocery delivery services. Autonomous systems have been demonstrably effective at efficiently moving things across dense urban environments. Should commercial-grade drone fleets become highly reliable and safe, high-risk groups of people could employ drones to pick up and deliver groceries. Using drones in this way could help limit foot traffic in indoor environments by both shoppers and delivery people, helping to reduce the spread of the virus. Recent proposals for drones to employ mass transportation to save energy for longer-distance commutes could also be highly feasible during times when there are fewer people taking mass transportation for fear of viral spread[18]. Currently, such drone delivery services are limited to participants in the Unmanned Aircraft Systems (UAS) Integration Pilot Program (IPP) by the Federal Aviation Administration (FAA) for commercial drone delivery.[19] The FAA is rightfully concerned about the assurance of commercial drones today, but limitations have restricted those wishing to research and test these capabilities. Without the ability for assurance mechanisms to be tested for drone delivery, society will never be able to realize the full benefits of these autonomous systems during the pandemic. There is currently an opportunity for the FAA to explore broader assurance testing of delivery drones by researchers and companies wishing to employ their services.

Indoor shopping and even apartment living have become a concern for many during the COVID-19 pandemic. The lack of widespread testing, especially in the United States, has exacerbated concerns because those who are infected may not know it. In the meantime, contract tracing is "not going well" according to Dr. Anthony Fauci, Director of the National Institute of Allergy and Infectious Disease (NIAID), a leading authority on the US COVID-19 response.[20] Existing highly-automated surveillance equipment in many buildings, such as infrared cameras, could be re- purposed to help with identifying those with fevers and facilitate contact tracing through facial recognition. Employing surveillance equipment in this way could be controversial, though, if steps are not taken to assure citizens of the security of the collected data and the privacy of those recorded by the autonomous systems. Means for securing and anonymizing such data are available, but not consistently deployed. Therefore, should society want to realize the benefits of autonomous systems during the pandemic, steps to enforce the consistent provisioning of security and privacy for surveillance equipment are essential.

---

[18] https://spectrum.ieee.org/automaton/robotics/drones/delivery-drones-could-hitchhike-on-public-transit-to-massively-expand-their-range

[19] https://www.faa.gov/uas/advanced_operations/package_delivery_drone/

[20] https://www.cnbc.com/2020/06/26/coronavirus-contact-tracing-is-not-going-well-fauci-says.html

Autonomous systems, as described, are no longer science fiction. However, their employment and benefit to society, especially in the context of smart cities, depends entirely on if they are assured. Smart cities are already transforming how people live and interact. Still, to reliably apply autonomous systems to use cases such as managing a pandemic environment, the safety and security of these systems are critical.

## 5.4 Mobility

Autonomous vehicles have the potential to improve the safety, efficiency, and sustainability of the transportation system. They would provide new mobility options, especially for those with some form of disability. Autonomous vehicles would take human driver error out of the equation and allow the time and money spent commuting to be put to better use (possibly at the expense of increased vulnerability to other sources of human error, e.g., in design, validation, and maintenance).

Assured autonomy plays a vital role in the performance of autonomous vehicles. In particular, the United States Code for Motor Vehicle Safety (Title 49, Chapter 301) defines motor vehicle safety as "the performance of a motor vehicle or motor vehicle equipment in a way that protects the public against unreasonable risk of accidents occurring because of the design, construction, or performance of a motor vehicle, and against unreasonable risk of death or injury in an accident." Additionally, the assurance of autonomous vehicles potentially may shape the public attitude toward the adoption of the technology, which is essential for its benefits to be realized.

As it turns out, the development timeline of autonomous vehicles is much longer than many early predictions.[21]

The uncertain and unstructured nature of environments in which the vehicles operate highly contributes to this delay as it adds complexity to the technical challenges discussed in Section 4. First, the open-world challenge leads to an unreasonably large number of test scenarios and drives the question of edge cases from the safety and verification as well as the certification and regulation aspects. Furthermore, the lack of system specifications and requirements is a significant impediment toward formal methods realizing their full potential in this domain.

Such requirements need to be precise and not to rely on assumptions regarding human nature or capabilities, including, for example, the maturity of judgment of 16-year-olds, as required in the case of human drivers. Finally, as edge cases potentially involve violation of some requirements, a hierarchy or an order of importance among the requirements needs to be defined. This potentially involves resolving the technical challenge related to ethics. How safe is safe enough to deploy this technology and what weight should be given to risks to different demographics and stakeholders, such as vulnerable road user safety vs. litigation exposure for developers?

In their early adoption, autonomous vehicles will likely share the road with human-driven vehicles. Additionally, they may need to hand over the control of the vehicle to the drivers under certain situations, especially those with less than level-5 autonomy. The challenge of human-system integration and trust needs to be addressed to ensure that the introduction of autonomous vehicles does not cause more accidents in the overall transportation system due to factors such as their conservative behaviors, inconsistencies with local driving practices, or confusion over the control of the vehicle. Security is also essential and is not an after-the-fact question since malicious attacks on sensor measurements or the vehicle's network may lead to fatal consequences. The current regulatory regime in the US is "self-certification," with regulatory oversight of potential computer-based system faults done primarily in reaction to on-road incidents and loss events. There is no consensus on how "driver licenses" will be handled for autonomous vehicles.

---

[21] https://spectrum.ieee.org/transportation/self-driving/surprise-2020-is-not-the-year-for-selfdriving-cars

The US DOT currently has a "non- regulatory" policy toward automated vehicle technology.[22]

## Environmental robotics and their impact on the future of work

Autonomous robots have the potential to address urgent environmental problems -- cleaning the oceans, recycling waste, and managing forests, for example. Their impact presents a complex composition of social, economic, and political challenges. For example, the recycling industry generates over $117B in economic activity in the US alone and employs more than half a million workers. Steady increases in domestic waste production, inefficiencies in current waste-handling practices, and changing international regulations have created an urgent need for innovative waste management strategies. While the industry heavily depends on human workers and manual work, recruiting and retaining workers for these stressful, hazardous, and poorly paid jobs have become challenging. Any shift in the industry hinges on reaching consensus among a large set of stakeholders–including consumers, packaging suppliers, recycling companies, workers, technology providers, and regulatory bodies – all with potentially conflicting interests.

While autonomous robots offer obvious opportunities in the waste management and recycling industry, their adoption for narrowly defined technical objectives, despite their short-term benefits, may result in unintended consequences. Therefore, an interdisciplinary, holistic approach is necessary. This approach must consider the implications of autonomous robots for the future of recycling work and workers and how such technology may change the characteristics and meaning of the job. For example, robots may reduce the frequency of injuries to waste-sorters., which currently is twice the industry average. Nevertheless, robots are currently capable of performing only those tasks that are also easiest and safest for human. Thus, the goal of realizing the most immediate efficiency and profitability gains from robotics conflicts with improving worker safety. A promising strategy for creating the intended levels of impact may be through properly designed human-autonomy co-work.

The COVID-19 pandemic has highlighted the importance of the waste management work as an essential service. It also has amplified the vulnerabilities of the recycling market. The crisis has impeded the international waste and materials trade, causing significant disruptions in the market. It has introduced additional risks for worker safety and new concerns for retaining workers and assuring work continuity. Therefore, an expedited adoption of autonomous robots may be expected in the near future. Such an adoption will increase the importance of assurance in the deployment and operation of autonomous robots and of a number of challenges discussed in the preceding sections, including safety, human-autonomy interactions, regulation, and possibly even ethics and privacy.

---

[22] https://www.nhtsa.gov/technology-innovation/automated-vehicles-safety

## 6. Workshop participants and contributors

The workshop organizers would like to thank the following workshop participants for their participation in the meetings. They would also like to thank Keith Marzullo (University of Maryland), Dierdre Mulligan (University of California Berkeley), David Mussington (University of Maryland), Fred Schneider (Cornell University), Berk Calli (Worcester Polytechnic Institute), and Ralph Wachter (NSF) for their help and guidance writing this report. Additional invaluable support was provided by the members of the NITRD Interagency Working Groups on Intelligent Robotics and Autonomy Systems and Computing Enabled Networked Physical Systems.

| First Name | Last Name | Affiliation | WS#1 | WS#2 | WS#3 |
|---|---|---|---|---|---|
| Behcet | Acikmese | University of Washington | | X | |
| Nisar | Ahmed | University of Colorado | X | | |
| Danette | Allen | National Aeronautics and Space Administration | X | | X |
| Julia | Badger | National Aeronautics and Space Administration | | X | |
| Sushil | Birla | U.S. Nuclear Regulatory Commission | X | | |
| Erik | Blasch | U.S. Air Force Research Lab | X | | |
| Nadya | Bliss | Arizona State University | X | X | X |
| Swarat | Chaudhuri | University of Texas at Austin | | X | X |
| Jessie | Chen | Army Research Laboratory | X | | |
| Erin | Chiou | Arizona State University | | X | |
| Darren | Cofer | Rockwell Collins | X | | X |
| Nancy | Cooke | Arizona State University | X | X | X |
| Missy | Cummings | Duke University | X | | |
| Anton | Dahbura | Johns Hopkins University Institute for Assured Autonomy | | X | |
| Jeremy | Daily | Colorado State University | | X | X |
| David | Danks | Carnegie Mellon University | | X | |
| Tom | Dietterich | Oregon State University | | X | |
| Warren | Dixon | University of Florida | X | | |
| Kevin | Dopart | U.S. Department of Transportation | X | | |
| Khari | Douglas | CRA/CCC | X | X | X |
| Katherine | Driggs-Campbell | University of Illinois | | X | |
| Gregory | Falco | Massachusetts Institute of Technology | | X | X |
| Michael | Fisher | University of Manchester, UK | | X | |
| Jake | Fries | National Coordination Office and the Networking and Information Technology Research and Development | X | | |
| Ryan | Gerdes | University of Virginia Tech | | X | |
| Mahsa | Ghasemi | University of Texas at Austin | X | X | |
| Julian | Goldman | Massachusetts General Hospital | X | | |
| Alwyn | Goodloe | National Aeronautics and Space Administration | X | X | X |
| Chris | Greer | National Institute of Standards and Technology | | | X |
| Keely | Griffith | Association for Unmanned Vehicle Systems International | | X | |
| Edward | Griffor | National Institute of Standards and Technology | X | | |
| Peter | Harsha | CRA/CCC | X | X | |
| David | Hendry | University of Washington | | X | X |
| Laura | Humphrey | Air Force Research Laboratory | X | | |

| First | Last | Affiliation | | | |
|---|---|---|---|---|---|
| Todd | Humphreys | The University of Texas at Austin | X | | |
| Matthew | Johnson | University of West Florida | X | | |
| Phil | Koopman | Carnegie Mellon University | X | | X |
| David | Kuehn | U.S. Department of Transportation | X | | X |
| Benjamin | Kuipers | University of Michigan | | X | |
| Cara | LaPointe | Johns Hopkins University Institute for Assured Autonomy | X | X | X |
| John | Launchbury | Galois | X | | |
| Craig | Lennon | Army Research Laboratory | X | | X |
| Frederick | Leve | Air Force | X | | |
| Ashley | Llorens | Johns Hopkins University Applied Physics Lab | X | X | X |
| Dan | Lopresti | Lehigh University | X | | X |
| Rahul | Mangharam | University of Pennsylvania | | X | X |
| William | Martin | National Security Agency | X | | |
| Katina | Michael | Arizona State University | | X | |
| Joel | Mozer | U.S. Air Force | X | | |
| Natasha | Neogi | National Aeronautics and Space Administration | X | | X |
| Tristan | Nguyen | U.S. Air Force | X | | |
| Meeko | Oishi | University of New Mexico | X | | |
| Hiro | Ono | National Aeronautics and Space Administration | X | | X |
| Miroslav | Pajic | Duke University | X | | |
| Sean | Phillips | Air Force Research Laboratory | X | | |
| Abraham | Puthuvana Vinod | University of Texas at Austin | | X | |
| Alexander | Ray | Open AI | | X | |
| Signe | Redfield | U.S. Naval Research Laboratory | X | | |
| Heather | Roff | Johns Hopkins University Applied Physics Lab | X | | X |
| Brian | Sadler | Army Research Laboratory | X | | |
| Ann | Schwartz | CRA/CCC | X | X | |
| Ben | Shneiderman | University of Maryland | | X | |
| Howie | Shrobe | Massachusetts Institute of Technology | X | X | X |
| Ankur | Srivastava | University of Maryland | X | | |
| Marc | Steinberg | Office of Naval Research | X | | |
| Paulo | Tabuada | University of California, Los Angeles | | X | |
| Ufuk | Topcu | University of Texas at Austin | X | X | X |
| Ralph | Wachter | National Science Foundation | X | | X |
| Albert | Wavering | National Institute of Standards and Technology | X | | |
| Sam | Weber | Office of Naval Research | | | X |
| Jamie | Winterton | Arizona State University | | X | |
| Tichakorn | Wongpiromsarn | Iowa State University | X | X | X |
| Helen | Wright | CRA/CCC | X | X | X |
| Mohammed | Yousuf | U.S. Department of Transportation | | X | |
| Lenore | Zuck | University of Illinois | X | X | X |

# NOTES

**NOTES**

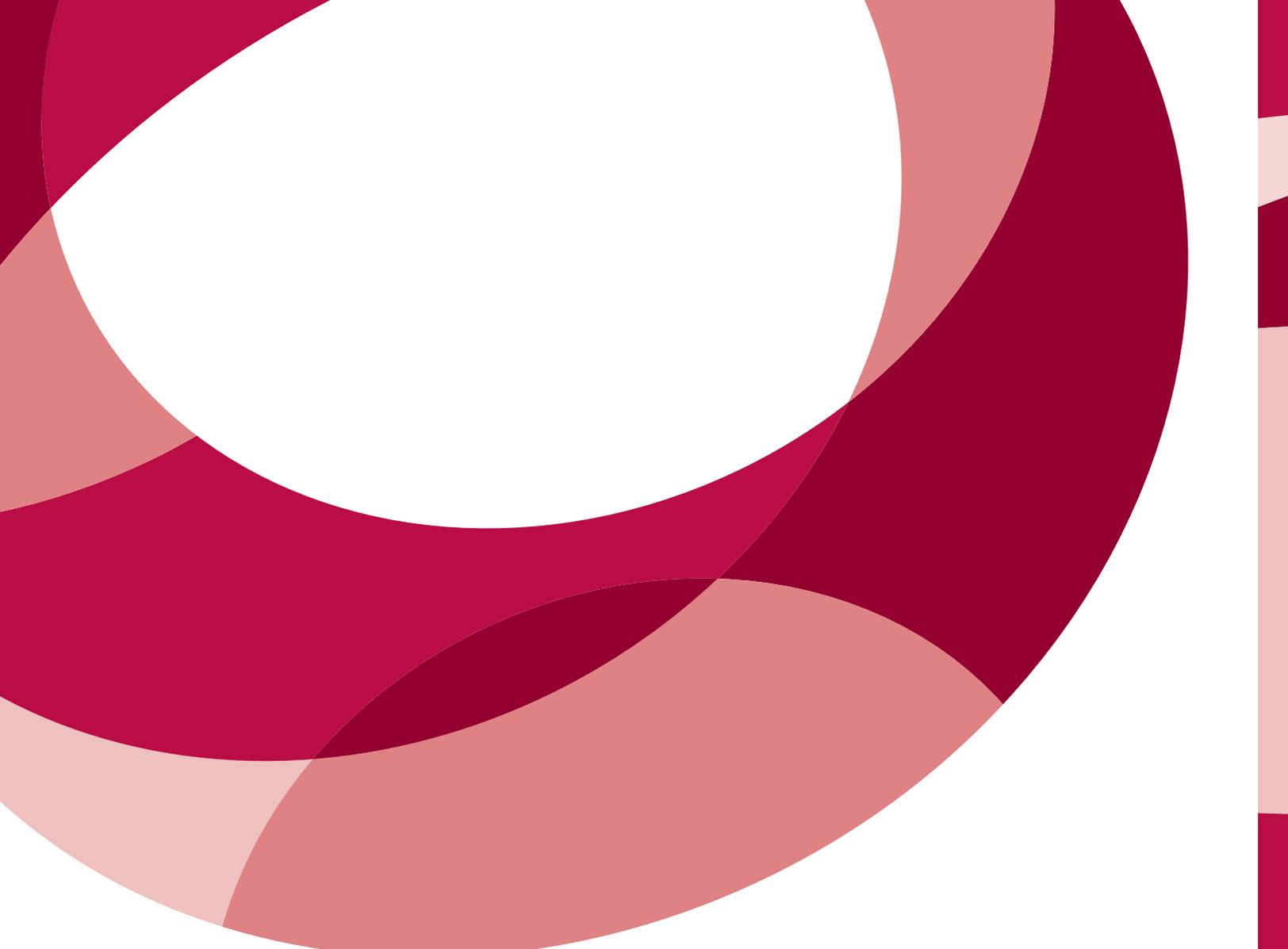

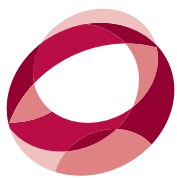

**CCC**
Computing Community Consortium
Catalyst

1828 L Street, NW, Suite 800
Washington, DC 20036
P: 202 234 2111 F: 202 667 1066
www.cra.org cccinfo@cra.org